\documentstyle[prl,aps,floats,psfig]{revtex}

\begin{document}
\twocolumn[
\hsize\textwidth\columnwidth\hsize\csname@twocolumnfalse\endcsname

\draft

\title{What happens to the quantum Hall effect when
magnetic-field-induced spin-density wave moves}
\author{Victor M. Yakovenko\cite{a} and Hsi-Sheng Goan\cite{b} }
\address{Department of Physics and Center for Superconductivity Research,
University of Maryland, College Park, MD 20742, USA}
\date{{\bf E-print cond-mat/9505016}, May 3, 1995 }
\maketitle

\begin{abstract}
   The influence of the motion of a magnetic-field-induced
spin-density wave (FISDW) on the quantum Hall effect in a
quasi-one-dimensional conductor is studied theoretically.  In the
ideal case of a free FISDW, it is found that the counterflow of the
FISDW precisely cancels the quantum Hall current, so the resultant
Hall conductivity is zero.  In real systems, the Hall conductivity
should vanish at the high frequencies, where the pinning and the
damping can be neglected, and the dynamics of the FISDW is dominated
by inertia.
\end{abstract}

\pacs{\it
To be published by World Scientific Publishing Co.\ in the Proceedings
of Physical Phenomena at High Magnetic Fields--II Conference,
Tallahassee, May 6--9, 1995 }

\bigskip

]

   It is known experimentally\cite{Cooper88} and understood
theoretically\cite{Poilblanc87,Yakovenko91} that the
magnetic-field-induced spin-density-wave (FISDW) state, observed in
the (TMTSF)$_2$X organic quasi-one-di\-men\-si\-o\-nal conductors,
exhibits the quantum Hall effect.  In the theoretical explanation of
this effect, it is assumed that the FISDW is pinned and acts on
electrons as a static potential.  The Hall conductivity, calculated in
the presence of this potential at zero temperature, is quantized.

   On the other hand, under certain conditions, the density wave in a
quasi-one-dimensional conductor can move\cite{Gruner88}.  It is
interesting to study how this motion would influence the quantum Hall
effect.  Since the density-wave condensate can move only along the
chains, at first sight, this purely one-dimensional motion cannot
contribute to the Hall effect which is essentially a two-dimensional
effect.  Nevertheless, it is shown below that in the case of the
FISDW, unlike in the case of a regular charge- or spin-density wave
(CDW/SDW), a {\it nonstationary} motion of the condensate does produce
a non-trivial contribution to the Hall conductivity.  This effect is
found at zero temperature in the absence of normal carriers and has
the same origin as the quantum Hall effect.  In the ideal system where
the FISDW is not pinned or damped, the contribution due to the FISDW
motion precisely cancels the bare quantum Hall term, so that {\it the
resultant Hall conductivity is zero}.  In real systems, this effect
should manifest itself at high enough frequencies where the dynamics
of the FISDW is dominated by inertia, and the pinning and the damping
can be neglected. On the other hand, the effect cannot be observed in
the DC measurements where the FISDW can be depinned by strong electric
field. In this paper, we present only a heuristic,
semiphenomenological outline, whereas a systematic derivation will be
given elsewhere. Some of these results were briefly reported also in
other conference proceedings\cite{Yakovenko93}.

   Let us consider a two-dimensional system where electrons are
confined to the chains parallel to the $x$-axis, and the spacing
between the chains along the $y$-axis is equal to $b$.  Magnetic field
$H$ is applied along the $z$-axis perpendicular to the
$(x,y)$-plane. The system is in the FISDW state at zero temperature.
Let us consider first the case where the electric field $E_y$ is
applied perpendicular to the chains. The electron Hamiltonian can be
written as
\begin{eqnarray}
{\cal H}&=&\frac{{\hbar}^2{k_x}^2}{2m} +
{\Delta_0}\cos(Q_x x+\Theta) \nonumber \\
&& +2t_b\cos(k_y b-q_x x +\omega_y t).
\label{Ham}
\end{eqnarray}
Here $\hbar=h/2\pi$ is the Planck constant, $m$ is the electron mass,
$k_x$ and $k_y$ are the electron wave vectors along and perpendicular
to the chains, and $t_b$ is the amplitude of tunneling between the
chains. In the gauge ${A_y}=Hx-c{E_y}t$ and $A_x=A_z=0$, the magnetic
and the transverse electric fields appear in the Hamiltonian
(\ref{Ham}) through the Peierls--Onsager substitution $k_y\rightarrow
k_y-eA_y/c\hbar$, so $q_x=ebH/\hbar c$ and
${\omega_y}=eb{E_y}/{\hbar}$ ($e$ is the electron charge and $c$ is
the velocity of light).  The FISDW potential is represented by the
second term in Eq.\ (\ref{Ham}) with the amplitude $\Delta_0$ and the
phase $\Theta$. It is well known\cite{Poilblanc87,Yakovenko91} that
the longitudinal wave vector of the FISDW is equal to
$Q_x=2{k_F}-Nq_x$, where $k_F$ is the Fermi wave vector and $N$ is an
integer. For simplicity, we set the transverse wave vector of the
FISDW to zero.

   We see that, in the presence of the magnetic field $H$, the hopping
term in Eq.\ (\ref{Ham}) acts as a potential, periodic along the
chains with the wave vector $q_x$ proportional to $H$. In the presence
of the transverse electric field $E_y$, this potential moves along the
chains with the velocity $\omega_y/q_x=cE_y/H$ proportional to
$E_y$. This velocity is nothing but the drift velocity in crossed
electric and magnetic fields. The FISDW potential may also move along
the chains\cite{Gruner88}, in which case its phase $\Theta$ depends on
time $t$, and the velocity of the motion is proportional to the time
derivative $\dot{\Theta}$. Since we are interested only in a spatially
homogeneous motion of the FISDW, we assume that $\Theta$ depends only
on $t$ and not on the coordinates $x$ and $y$. We also assume that
both potentials move very slowly, adiabatically, which is the case
when the electric field is sufficiently weak.

   Now, we are going to calculate the current along the chains
produced by the motion of the potentials. Since there is an energy gap
at the Fermi level, following the arguments of
Laughlin\cite{Laughlin81} we can say that an integer number of
electrons $N_1$ is transferred from one end of a chain to another when
the FISDW potential shifts by its period $L_1=2\pi/Q_x$. The same is
true for the motion of the hopping potential with an integer $N_2$ and
the period $L_2=2\pi/q_x$. Suppose that the first potential shifts by
an infinitesimal displacement $dx_1$ and the second by $dx_2$. The
total transferred charge $dq$ would be the sum of the prorated amounts
of $N_1$ and $N_2$:
\begin{equation}
dq=eN_1\frac{dx_1}{L_1}+eN_2\frac{dx_2}{L_2}.
\label{dq}
\end{equation}
Now, suppose that both potentials are shifted by the same displacement
$dx=dx_1=dx_2$. In this case, we can also write that
\begin{equation}
dq=e\rho\,dx,
\label{rho}
\end{equation}
where $\rho=4k_F/2\pi$ is the concentration of electrons. Equating
(\ref{dq}) and (\ref{rho}) and substituting the expressions for
$\rho$, $L_1$, and $L_2$, we find the following Diophantine-type
equation\cite{Zak}:
\begin{equation}
4k_F=N_1 (2 k_F-Nq_x)+N_2 q_x.
\label{Diophant}
\end{equation}
Since $k_F/q_x$ is, in general, an irrational number, the only
solution of Eq.\ (\ref{Diophant}) for the integer $N_1$ and $N_2$ is
$N_1=2$ and $N_2=N_1 N=2N$.

   Dividing Eq.\ (\ref{dq}) by the time increment $dt$ and the
distance between the chains $b$, we find the density of current along
the chains, $j_x$. Taking into account that according to Eq.\
(\ref{Ham}) the displacements of the potentials are related to their
phases: $dx_1=-d\Theta/Q_x$ and $dx_2=\omega_ydt/q_x$, we find the
final expression for $j_x$:
\begin{equation}
j_x=-\frac{e}{\pi b}\dot{\Theta} + \frac{2Ne^2}{h}E_y.
\label{jx}
\end{equation}
The first term in Eq.\ (\ref{jx}) represents the contribution of the
FISDW motion, the so-called Fr\"{o}hlich conductivity\cite{Gruner88}.
The second term describes the quantum Hall
effect\cite{Poilblanc87,Yakovenko91}. The integer number $N$ in the
quantized Hall conductivity $\sigma_{xy}=2Ne^2/h$ is the same as that
in the FISDW wave vector $Q_x=2k_F-Nq_x$.

   To complete solution of the problem, it is necessary to find how
$\dot{\Theta}$ depends on $E_y$. For this purpose, we need the
equation of motion of $\Theta$, which can be derived once we know the
Lagrangian of the system, $L$.  Two terms in $L$ can be readily
recovered taking into account that the current density $j_x$, given by
Eq.\ (\ref{jx}), is the variational derivative of the Lagrangian with
respect to the electromagnetic vector-potential $A_x$: $j_x=c\,\delta
L/\delta A_x$.  Written in the gauge-invariant form, the recovered
part of the Lagrangian is equal to
\begin{equation}
L_1=-\sum_{i,j,k}\frac{Ne^2}{2\pi\hbar c}\varepsilon_{ijk}A_iF_{jk}
-\frac{e}{\pi b}\Theta E_x,
\label{L1}
\end{equation}
where $\varepsilon_{ijk}$ is the antisymmetric tensor with the indices
$i,j,k=t,x,y$; $A_i$ and $F_{jk}$ are the vector-potential and the
tensor of the electromagnetic field, and $E_x\equiv F_{tx}$ is the
electric field along the chains.  The first term in Eq.\ (\ref{L1}) is
the so-called Chern--Simons term responsible for the quantum Hall
effect\cite{Yakovenko91}.  The second term describes the interaction
of the density-wave condensate with the electric field along the
chains\cite{Gruner88}.

   Lagrangian (\ref{L1}) should be supplemented with the kinetic
energy of the FISDW condensate, $K$.  The FISDW potential itself has
no inertia because it is produced by the instantaneous Coulomb
interaction between electrons, so $K$ originates completely from the
kinetic energy of the electrons which are confined under the FISDW
gap.  The latter energy is proportional to the square of their average
velocity, which, in turn, is proportional to the electric current
along the chains:
\begin{equation}
K=\frac{\pi\hbar b}{4v_Fe^2}\,j_x^2,                  \label{K}
\end{equation}
where $v_F$ is the Fermi velocity. Substituting Eq.\ (\ref{jx}) into
Eq.\ (\ref{K}), expanding, and omitting an unimportant term proportional
to $E_y^2$, we obtain the second part of the Lagrangian of the system:
\begin{equation}
L_2=\frac{\hbar}{4\pi bv_F}\dot{\Theta}^2
-\frac{eN}{2\pi v_F}\dot{\Theta}E_y.                \label{L2}
\end{equation}
The first term in Eq.\ (\ref{L2}) is the same as the kinetic energy of
a purely one-dimensional density wave\cite{Gruner88} and is not
specific to the FISDW.  The most important is the second term which
describes the interaction of the FISDW motion and the electric field
perpendicular to the chains. This term is allowed by symmetry in the
considered system and has the structure of a mixed vector--scalar
product:
\begin{equation}
{\bf v}\,[{\bf E}\times{\bf H}].
\label{vEH}
\end{equation}
Here, ${\bf v}$ is the velocity of the FISDW which is proportional to
$\dot{\Theta}$ and is directed along the chains, that is, along the
$x$-axis.  The magnetic field ${\bf H}$ is directed along the
$z$-axis, thus, allowing the electric field ${\bf E}$ to enter only
through the component $E_y$.  Comparing formula (\ref{vEH}) with the
second term in Eq.\ (\ref{L2}), one should take into account that the
magnetic field enters the second term implicitly, through the integer
$N$, which depends on $H$ and changes sign when $H$ changes sign.

   Varying the total Lagrangian $L=L_1+L_2$, given by Eqs.\ (\ref{L1})
and (\ref{L2}), with respect to $\Theta$, we find the equation of motion
of $\Theta$:
\begin{equation}
\ddot{\Theta}=-\frac{2ev_F}{\hbar}E_x + \frac{eNb}{\hbar}\dot{E_y}.
\label{EOM}
\end{equation}
In Eq.\ (\ref{EOM}), the first two terms constitute the standard
one-dimensional equation of motion of the density wave\cite{Gruner88},
whereas the last term, proportional to the time derivative of $E_y$,
which originated from the second term in Eq.\ (\ref{L2}), describes
the influence of the electric field across the chains on the motion of
the FISDW.

   Taking into account that $E_x=0$ and integrating Eq.\ (\ref{EOM}),
we find that $\dot{\Theta}=eNbE_y/\hbar$; thus, the second term in
Eq.\ (\ref{jx}) (the Fr\"{o}hlich conductivity of the FISDW) precisely
cancels the first term (the quantum Hall current), so {\it the
resulting Hall current is equal to zero}.  This result could have been
obtained without calculations by taking into account that the time
dependence $\Theta(t)$ is determined by the principle of minimal
action.  The relevant part of the action is given, in this case, by
Eq.\ (\ref{K}) which attains the minimal value when $j_x=0$.

   It is instructive to see how the nullification of the Hall
conductivity takes place in the case where the electric field is
directed along the chains.  Varying $L$ (Eqs.\ (\ref{L1}) and
(\ref{L2})) with respect to $A_y$, we find the density of current
perpendicular to the chains:
\begin{equation}
j_y=-\frac{2Ne^2}{h}E_x-\frac{eN}{2\pi v_F}\ddot{\Theta}.
\label{jy}
\end{equation}
In Eq.\ (\ref{jy}), the first term describes the quantum Hall
current, whereas the second term, proportional to the {\it acceleration}
of the FISDW condensate, comes from the second term in Eq.\  (\ref{L2})
and reflects the contribution of the FISDW motion along the chains to
the electric current across the chains.  According to the equation of
motion (\ref{EOM}), the electric field along the chains accelerates
the density wave: $\ddot{\Theta}=-2ev_FE_x/\hbar$, thus, the Hall
current (\ref{jy}) vanishes.

   It is clear, however, that in stationary, DC measurements the
acceleration of the FISDW, discussed in the previous paragraph, cannot
last forever.  Any friction or dissipation will inevitably stabilize
the motion of the density wave to a steady flow with zero
acceleration.  In this steady state, the second term in Eq.\
(\ref{jy}) vanishes, and the current $j_y$ recovers its quantum Hall
value.  The same is true in the case where the electric field is
perpendicular to the chains.  In that case, the dissipation eventually
stops the motion of the FISDW along the chains and restores $j_x$,
given by Eq.\ (\ref{jx}), to the quantum Hall value. The conclusion is
that the contribution of the moving FISDW condensate to the Hall
conductivity is essentially nonstationary and cannot be observed in DC
measurements.

\begin{figure}
\centerline{\psfig{file=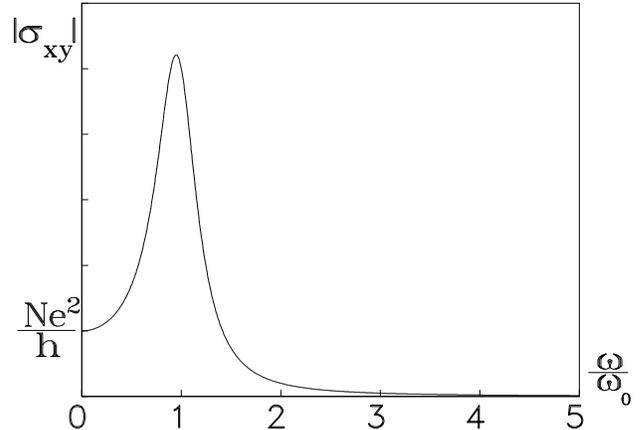,width=\linewidth,angle=-90}}
\caption{\protect\footnotesize\baselineskip=13pt
Hall conductivity of the FISDW system as a function of frequency
$\omega$ normalized to the pinning frequency $\omega_0$.}
\label{fig}
\end{figure}

   On the other hand, the effect can be seen in AC experiments.  To be
realistic, let us add damping and pinning\cite{Gruner88} to the
equation of motion of the FISDW (\ref{EOM}):
\begin{equation}
\ddot{\Theta}+\frac{1}{\tau}\dot{\Theta}+\omega_0^2\Theta
=-\frac{2ev_F}{\hbar}E_x + \frac{eNb}{\hbar}\dot{E_y},
\label{fric}
\end{equation}
where $\tau$ is the relaxation time and $\omega_0$ is the pinning
frequency.  Solving Eq.\ (\ref{fric}) via the Fourier transformation
from the time $t$ to the frequency $\omega$ and substituting the
result into Eqs.\ (\ref{jx}) and (\ref{jy}), we find the Hall
conductivity as a function of the frequency:
\begin{equation}
\sigma_{xy}(\omega)=\frac{2Ne^2}{h}\frac{\omega_0^2-i\omega/\tau}
{\omega_0^2-\omega^2-i\omega/\tau}.
\label{omega}
\end{equation}
The absolute value of the Hall conductivity, $|\sigma_{xy}|$, computed
from Eq.\ (\ref{omega}) is plotted in the Fig.\ \ref{fig} as a
function of $\omega/\omega_0$ for $\omega_0\tau=2$.  As we can see in
the Figure, the Hall conductivity is quantized at zero frequency and
has a resonance at the pinning frequency.  At the higher frequencies,
where the pinning and the damping can be neglected and the system
effectively behaves as an ideal, purely inertial system considered
above, the Hall conductivity does decrease toward zero.

   The frequency dependence of the Hall conductivity in regular
semiconductor quantum Hall systems was measured using the technique of
crossed wave guides\cite{Kuchar,Galchenkov}. Unfortunately, no such
measurements were performed in the FISDW systems. These measurements
would be extremely interesting. To give a crude estimate of the
required frequency range, we quote the value of the pinning frequency
$\omega_0\sim$ 3 GHz $\sim$ 0.1 K $\sim$ 10 cm for a regular (not
magnetic-field-induced) SDW in (TMTSF)$_2$PF$_6$\cite{Quinlivan}.

   This work was partially supported by NSF under Grant DMR--9417451
and by A.~P.~Sloan Foundation.

\end{document}